\documentclass[twocolumn]{aastex62}

\newcommand{\acronym}[1]{{\small{#1}}}
\newcommand{\package}[1]{\textsl{#1}}
\newcommand{\Gaia}{\textsl{Gaia}}
\newcommand{\gaia}{\textsl{Gaia}}
\newcommand{\hst}{\textsl{HST}}

\newcommand{\lcdm}{\acronym{$\Lambda$CDM}}

\newcommand{\changes}[1]{#1}

\shorttitle{dynamical evidence of a dark halo substructure}
\shortauthors{bonaca et al.}

\begin{document}\sloppy\sloppypar\raggedbottom\frenchspacing

\title{\textbf{%
The Spur and the Gap in GD-1:\\
Dynamical evidence for a dark substructure in the Milky Way halo
}}

\correspondingauthor{Ana Bonaca}
\email{ana.bonaca@cfa.harvard.edu}

\author[0000-0002-7846-9787]{Ana Bonaca}
\affil{Harvard--Smithsonian Center for Astrophysics, 60 Garden St, Cambridge, MA 02138, USA}

\author[0000-0003-2866-9403]{David W. Hogg}
\affil{Center for Cosmology and Particle Physics, Department of Physics, New York University, 726~Broadway, New York, NY 10003, USA}
\affil{Center for Data Science, New York University, 60 Fifth Ave, New York, NY 10011, USA}
\affil{Max-Planck-Institut f\"ur Astronomie, K\"onigstuhl 17, D-69117 Heidelberg}
\affil{Flatiron Institute, 162 Fifth Ave, New York, NY 10010, USA}

\author[0000-0003-0872-7098]{Adrian~M.~Price-Whelan}
\affil{Department of Astrophysical Sciences, Princeton University, Princeton, NJ 08544, USA}

\author[0000-0002-1590-8551]{Charlie Conroy}
\affil{Harvard--Smithsonian Center for Astrophysics, 60 Garden St, Cambridge, MA 02138, USA}

\begin{abstract}\noindent
We present a model for the interaction of the GD-1 stellar stream with a massive perturber that naturally explains many of the observed stream features, including a gap and an off-stream spur of stars.
The model involves an impulse by a fast encounter, after which the stream grows a loop of stars at different orbital energies.
At specific viewing angles, this loop appears offset from the stream track.
A quantitative comparison of the spur and gap features prefers models where the perturber is in the mass range of $10^6\,\rm M_\odot$ to $10^8\,\rm M_\odot$.
Orbit integrations back in time show that the stream encounter could not have been caused by any known globular cluster or dwarf galaxy \changes{with a determined orbit}, and mass, size and impact-parameter arguments show that it could not have been caused by a molecular cloud in the Milky Way disk.
The most plausible explanation for the gap-and-spur structure is an encounter with a dark-matter substructure, like those predicted to populate galactic halos in \lcdm\ cosmology.
However, the expected densities of \lcdm\ subhalos in this mass range and in this part of the Milky Way are $2-3\,\sigma$ lower than the inferred density of the GD-1 perturber.
This observation opens up the possibility that detailed observations of streams could measure the mass spectrum of dark-matter substructures and even identify individual substructures and their orbits in the Galactic halo.
\end{abstract}


\section{Introduction}
\label{sec:intro}
In the currently preferred \lcdm\ cosmological model of cold dark matter (CDM) with dark energy ($\Lambda$), dark matter forms clumps of very low masses \citep[e.g.,][]{springel2008}.
Through mergers, these clumps grow to become massive halos that include a number of distinct lower mass clumps, or subhalos \citep[e.g.,][]{whiterees1978}.
Baryons can only be retained in halos more massive than $\sim10^8-10^9\,\rm M_\odot$ \citep[e.g.,][]{efstathiou1992, bullock2000}, which agrees well with the lowest-mass galaxies discovered around the Milky Way \citep[e.g.,][]{sg2007, martin2008}.
A critical prediction of the CDM paradigm is the existence of dark subhalos less massive than $\lesssim10^8\,\rm M_\odot$.

Alternative cosmological models have been proposed that behave like \lcdm\ on large scales, but have less structure on small scales.
In the case of warm dark matter \citep[e.g.,][]{bode2001}, this is accomplished with a dark matter particle that is less massive ($m\sim\rm\, keV$) than the cold dark matter particle ($m\gtrsim10\rm\, GeV$), and thus streams out of the lowest mass clumps.
The fuzzy dark matter model \citep[e.g.,][]{hu2000} posits an ultra-light dark matter particle ($m\sim10^{-22}\rm\, eV$) that exhibits quantum behavior on macroscopic scales, which prevents collapse of halos less massive than $\sim10^7\,\rm M_\odot$.
Therefore, a ruling on the existence of low-mass \changes{$(\lesssim10^7\,\rm M_\odot)$} dark matter subhalos would place strong constraints on the nature of dark matter \citep[e.g.,][]{bullockmbk2017, buckleypeter2017}\changes{, which can be further refined by measuring the minimum halo mass \citep[e.g.,][]{schmid1999, hofmann2001, loebzaldarriaga2005}}.

Albeit dark, low-mass subhalos should still exert gravitational influence, \changes{for example, fluctuations of the gravitational tidal field are sensitive to the presence of subhalos down to $10^{-6}\rm M_\odot$ \citep{penarrubia2018}.}
In a cosmological \changes{volume, however}, gravitational lensing is our most sensitive method of detecting gravitational perturbations.
And indeed, some strongly lensed galaxies require the presence of a subhalo in the lens plane to fully explain the distribution of light in the lensed system.
To date, subhalos in the mass regime $10^8-10^9\,\rm M_\odot$ have been identified in a number of lenses \citep[e.g.,][]{vegetti2012,hezaveh2016}.
However, these objects are expected to host stars (although at luminosities below the current detection threshold), so the search for lower mass, and truly dark, subhalos continues, primarily by increasing the sample size of analyzed lenses.

In the local universe, dynamically cold stellar streams are promising devices for measuring detailed properties of the matter distribution \citep[e.g.,][]{johnston1999, bh2018}.
Formed by stars escaping a disrupting globular cluster at slightly offset orbital energies, stellar streams are approximately one-dimensional structures in the 6D phase space.
An encounter between a stellar stream and a dark matter subhalo would perturb the orderly structure of the stellar stream, produce a gap in the stream density \citep[e.g.,][]{johnston2002, ibata2002, carlberg2012}, and, depending on the impact geometry, possibly also fold a part of the stream \citep[e.g.,][]{yoon2011}.
More than 40 thin stellar streams have been discovered in the Milky Way halo \citep{gc2016}, and the most prominent ones have already been searched for evidence of density variations.
The abundance of dark matter subhalos down to $\sim10^6\,\rm M_\odot$ inferred from the power-spectrum of density variations in streams is consistent with the \lcdm\ predictions \citep[e.g.,][]{carlberg2012,cg2013}.
In addition to dark matter subhalos, a number of physical processes and observational effects can alter stream morphology at this level \citep[e.g.,][]{kupper2008, amorisco2016, ibata2016}.
As a result, no stream has definitively established the presence of dark matter substructure.

So far, studies of stream gaps have been statistical in nature \citep[e.g.,][]{erkal2017}, mainly because the data were not good enough to identify individual gaps at high confidence.
However, thanks to the recently released \gaia\ data \citep{gdr2}, identification of stream member stars has become extremely efficient \citep[e.g.,][]{malhan2018}.
In turn, this has enabled the discovery of a perturbation site in the GD-1 stream \citep{pwb} -- a prime candidate for dark matter influence on a stellar stream.

In this work, we follow up this initial discovery with the first in-depth analysis of a perturbed stellar stream.
We first review the observed properties of GD-1 (\S\,\ref{sec:data}), and then develop a fiducial model of a perturbed stream that qualitatively matches GD-1 observations (\S\,\ref{sec:model}).
Next, we explore the range of impact parameters allowed by the spatial distribution of GD-1 (\S\,\ref{sec:perturber_properties}), and make predictions for its kinematics (\S\,\ref{sec:kinematics}).
Finally, we discuss the limitations of the current modeling framework (\S\,\ref{sec:caveats}), possible origin of the perturber (\S\,\ref{sec:origin}), and suggest strategies to distinguish between different origin scenarios (\S\,\ref{sec:future}).

\section{Observed features of the GD-1 stellar stream}
\label{sec:data}
GD-1 is the longest ($>100^\circ$, 10\,kpc) thin ($\sigma\approx12'$, 30\,pc) stellar stream discovered in the Galactic halo \citep{gd2006}.
Based on its width and length, GD-1 is expected to be extremely informative about the global distribution of matter in the Galaxy \citep{lux2013, bh2018}.
Indeed, dynamical modeling of GD-1 individually, and in concert with the tidal tails of the Palomar~5 globular cluster, has already revealed that the inner dark matter halo is spherical \citep{koposov2010, bowden2015, bovy2016}.
GD-1 data from the Sloan Digital Sky Survey \citep[SDSS,][]{york2000} has been analyzed for signatures of density variations \citep{cg2013}, but the contamination from the foreground Milky Way stars was too high to unambiguously attribute detected gaps to the stream itself.

\begin{figure*}
\begin{center}
\includegraphics[width=0.9\textwidth]{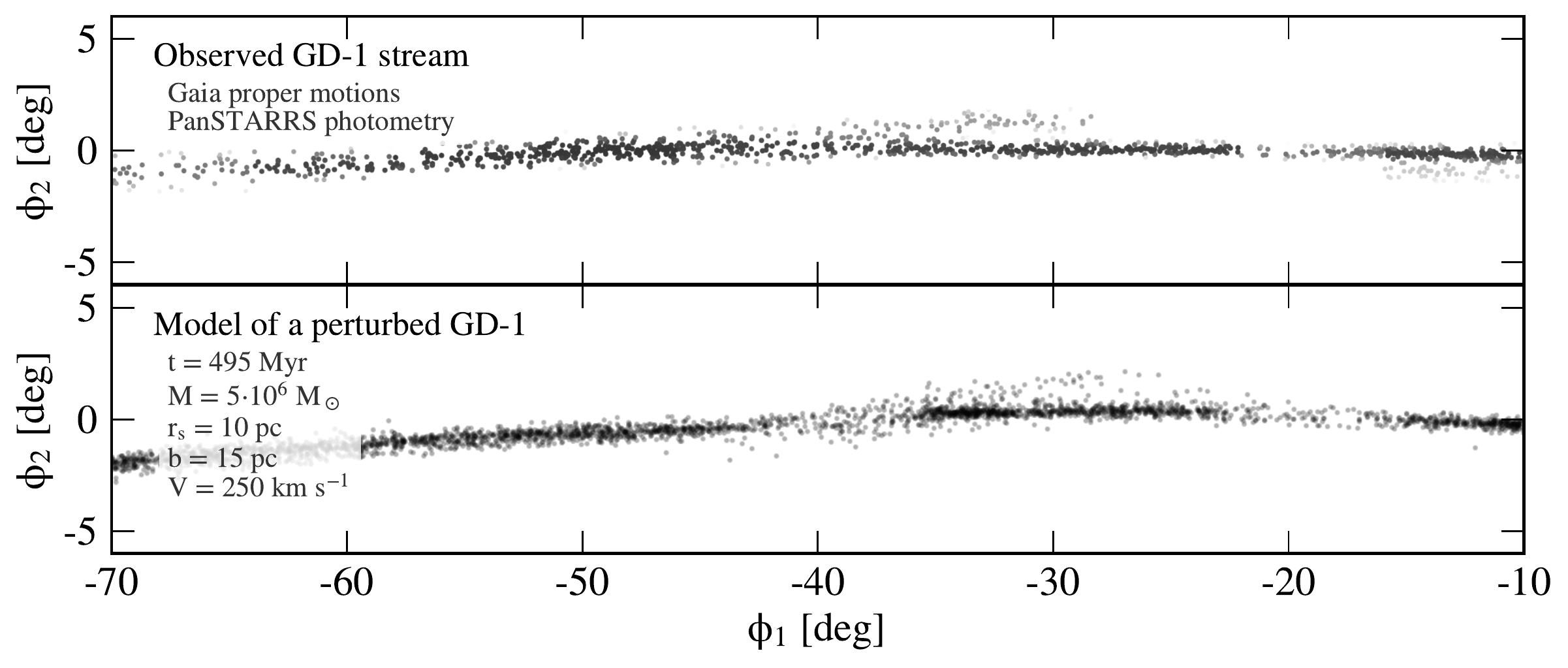}
\end{center}
\caption{(Top) Likely members of the GD-1 stellar stream, cleanly selected using Gaia proper motions and PanSTARRS photometry, reveal two significant gaps located at $\phi_1\approx-20^\circ$ and $\phi_1\approx-40^\circ$, and dubbed G-20 and G-40, respectively.
There is a long, thin spur extending for $\approx10^\circ$ from the G-40 gap.
(Bottom) An idealized model of GD-1, whose progenitor disrupted at $\phi_1\approx-20^\circ$ to produce the G-20 gap, and which has been perturbed by a compact, massive object to produce the G-40 gap.
The orbital structure of stars closest to the passing perturber is distorted into a loop of stars that after 0.5\,Gyr appears as an underdensity coinciding with the observed gap, and extends out of the stream similar to the observed spur.
}
\label{fig:fiducial}
\end{figure*}

Recently, \citet{pwb} used proper motions from the \gaia\ mission \citep{gdr2} and photometry from PanSTARRS \citep{ps} to confidently separate GD-1 stars from the Milky Way field stars.
This contamination-free view of GD-1 enabled the first unambiguous detection of gaps in a stellar stream, which remain evident in deeper imaging \citep{deboer2018}.
Additionally, the combined astrometric and photometric selection reveals GD-1 stars offset from the main stream track: an extended spur at $(\phi_1, \phi_2)\approx(-33^\circ,1^\circ)$ and a diffuse blob at $(\phi_1, \phi_2)\approx(-14^\circ,-1^\circ)$ in GD-1 coordinates (Figure~\ref{fig:fiducial}, top).
Patterns imparted by the complex selection function, confusion by background galaxies, or foreground dust do not coincide with these GD-1 features; instead, they are inherent properties of the stream itself \citep{pwb}.

To highlight the complex structure of the GD-1 stream, we present the distribution of likely stream members at the top of Figure~\ref{fig:fiducial}.
As a first step in finding likely members, we followed \citet{pwb} in selecting stars consistent with an old and metal-poor population at a distance of 8\,kpc, and moving retrograde with respect to the Galactic disk, with proper motions in the GD-1 reference frame $(\mu_{\phi_1}, \mu_{\phi_2})\approx(-7,0)\;\rm mas\,yr^{-1}$.
The spatial distribution of these stars in the $\phi_2$ direction (i.e. perpendicular to the stream) is modeled as a combination of a constant background, a stream component at the location of the main stream track, and one additional Gaussian component on either side of the main stream to capture stream features beyond the main track.
We solved for the normalization, position and width of every component by exploring the parameter space with an ensemble MCMC sampler \citep{Foreman-Mackey:2013}.
We used 256 walkers that ran for a total of 1280 steps, and kept the final 256 steps to generate posterior samples in these parameters.
The above procedure is a full-stream generalization of the calculation in \citep{pwb} that quantified the fraction of stars in the additional components at the locations of the spur and the blob.
Finally, we define a stream membership probability, $p_{mem}$, as the joint probability of a star belonging either to the main stream or the additional feature, evaluate these probabilities using MCMC samples and apply them to every star.
The upper panel of Figure~\ref{fig:fiducial} shows stars with $p_{mem}>0.5$, with larger and darker points representing stars with a higher membership probability.

Most likely GD-1 members trace a thin stream, whose width varies between $\sigma\approx10'$ and $30'$.
As noted by \citet{pwb}, the stellar density along the stream is not uniform, and there are two significant underdensities, or gaps, located at $\phi_1\approx-40^\circ$ and $\phi_1\approx-20^\circ$, which we refer to as G-40 and G-20, respectively.
The main focus of this work are structures related to the G-40 gap, so if not specified, the gap refers to G-40.
The additional, feature components are above the background density in the spur region, $\phi_1\approx-35^\circ$, and the blob region, $\phi_1\approx-15^\circ$, and consistent with zero along the rest of the stream.
In the following section we present a model of GD-1 that simultaneously explains the gap in the stream and the spur extending from the stream.

\section{Modeling the perturbed GD-1 stream}
\subsection{Setup and the fiducial model}
\label{sec:model}
Unlike the observed GD-1, a globular cluster disrupting on the GD-1 orbit in a simple --- analytic and smooth --- galaxy creates a stream that is also smooth \citep{pwb}.
This model follows stars as they leave the progenitor, and accounts for their epicylic motion relative to the progenitor's orbit \citep{kupper2008, kupper2010, fardal2015}.
The resulting pattern of over- and underdensities is much more uniform than the observed stream, so the full extent of density variations in GD-1 cannot be simply explained by the process of globular cluster disruption alone.
As inhomogeneities can also be introduced into a stellar stream by adding a perturbation to the gravitational potential \citep[e.g.,][]{sgv2008}, in this Section we present a model of the GD-1 stream that had a recent, close encounter with a dense, massive object.

As a first step in creating a model of the GD-1 stream, we follow \citet{pwb} in finding the orbit of the GD-1 progenitor by fitting the six-dimensional phase-space distribution of GD-1 stars.
We assume a spherical logarithmic potential with a circular velocity of 225\,km\,s$^{-1}$ for orbit integration.
This simple gravitational potential is very close to the best-fit model of GD-1 \citep{koposov2010, bowden2015}, and it also allows much faster force evaluations than the standard, multi-component model of the Milky Way.
\changes{The present-day location of the GD-1 progenitor is not firmly established \citep[e.g.,][]{pwb, webb2019}, so we adopted a scenario in which the progenitor is completely dissolved because it qualitatively reproduces a number of global GD-1 features \citep[see][]{pwb}.}

We assume that the GD-1 progenitor was a globular cluster of initial mass $7\times10^4\rm\,M_\odot$.
In our model, it started losing stars through evaporation 3\,Gyr ago and completely disrupted 300\,Myr before the present day.
We follow the progenitor's dissolution by releasing test particles from its Lagrange points (the mean ejection radius is $\sim150\,\rm pc$ initially), and produce a model of the stream following \citet{fardal2015}.
\changes{In this approach, the tidal radius from which the stars are being ejected is a function of the progenitor's mass and its position in the Galaxy.}
Although idealized, such models capture detailed properties of the more realistic, N-body, simulations of disrupting globular clusters \citep{kupper2012}.
The present day distribution of test particles is shown in GD-1 coordinates in the bottom of Figure~\ref{fig:fiducial}.
Had the progenitor survived to the present, it would be located at $\phi_1=-20^\circ$.
Instead, this model has a gap at that location, which coincides with the G-20 gap observed in GD-1.
The progenitor's initial mass and time of disruption were chosen to reproduce the stream width and the morphology of the more depleted observed gap.

However, the observed GD-1 stream has two prominent gaps.
To produce a model stream that also has a gap coinciding with the observed G-40 gap, our model also includes a massive and dense object on an orbit that crosses GD-1.
The parameters of the encounter were chosen to reproduce the observed morphology of the gap-and-spur feature in GD-1.
The perturber is represented by a \citet{hernquist1990} sphere of mass $5\times10^6\rm\,M_\odot$ and scale radius 10\,pc.
Its closest approach to GD-1 happened 495\,Myr ago with a relative distance of 15\,pc, which is smaller than the stream width.
The perturber came closest to GD-1 stars that are presently at $\phi_1\approx-38^\circ$.
During the encounter, nearby stars received a velocity kick from the perturber, and started moving towards the location of its closest approach.
In case of a weaker perturbation, e.g., one produced by a more diffuse perturber, the most significant component of the velocity kick is along the stream, which changes the orbital period of affected stars \citep{eb2015}.
On one side of the perturber, the affected stars have shorter orbital periods and hence speed by the unaffected stars, while on the other side they take longer to orbit the Galaxy, and lag behind the unaffected stream stars.
This creates a gap at the projected location of the closest approach, with a pile-up of stars on either side of the gap creating a signature double-horned profile \citep{carlberg2012}.
However, the perturber in our model is dense, so it imparts a significant velocity kick perpendicular to the stream as well as along the stream.
This leads to a loop of stars straying beyond the unperturbed stream track.
At the present, this loop is viewed nearly edge-on and looks like the observed spur (Figure~\ref{fig:fiducial}, bottom).

The stream model in the bottom of Figure~\ref{fig:fiducial} qualitatively matches many features in the observed GD-1 stream (Figure~\ref{fig:fiducial}, top).
Not only are both of the most prominent gaps reproduced at the right location and with the right size, but their density contrast is matched as well.
The G-20 gap, modeled as a disrupted progenitor, is almost completely depleted, while the G-40 gap, the location of the impact, still retains some stars.
Furthermore, the model features a spur of the correct offset from the main stream and correct length.
It is not a perfect model, for example, the model stream extends past the observed extent of GD-1.
Still, this model is a remarkably realistic rendition of the observed GD-1, so we next quantitatively explore the range of impact parameters that produce a good match to the observed stream.

\subsection{Constraining the GD-1 perturber}
\label{sec:perturber_properties}
In our fiducial model of the GD-1 stellar stream (Section~\ref{sec:model}), the encounter with a perturber introduced a gap in the stream and ejected a spur of GD-1 stars beyond the main stream.
In this section, we constrain the range in perturber's mass and size, its impact parameter, velocity, and time of perturbation that reproduce well the location and width of the gap, and the location and extent of the spur.

\begin{figure*}
\begin{center}
\includegraphics[width=0.9\textwidth]{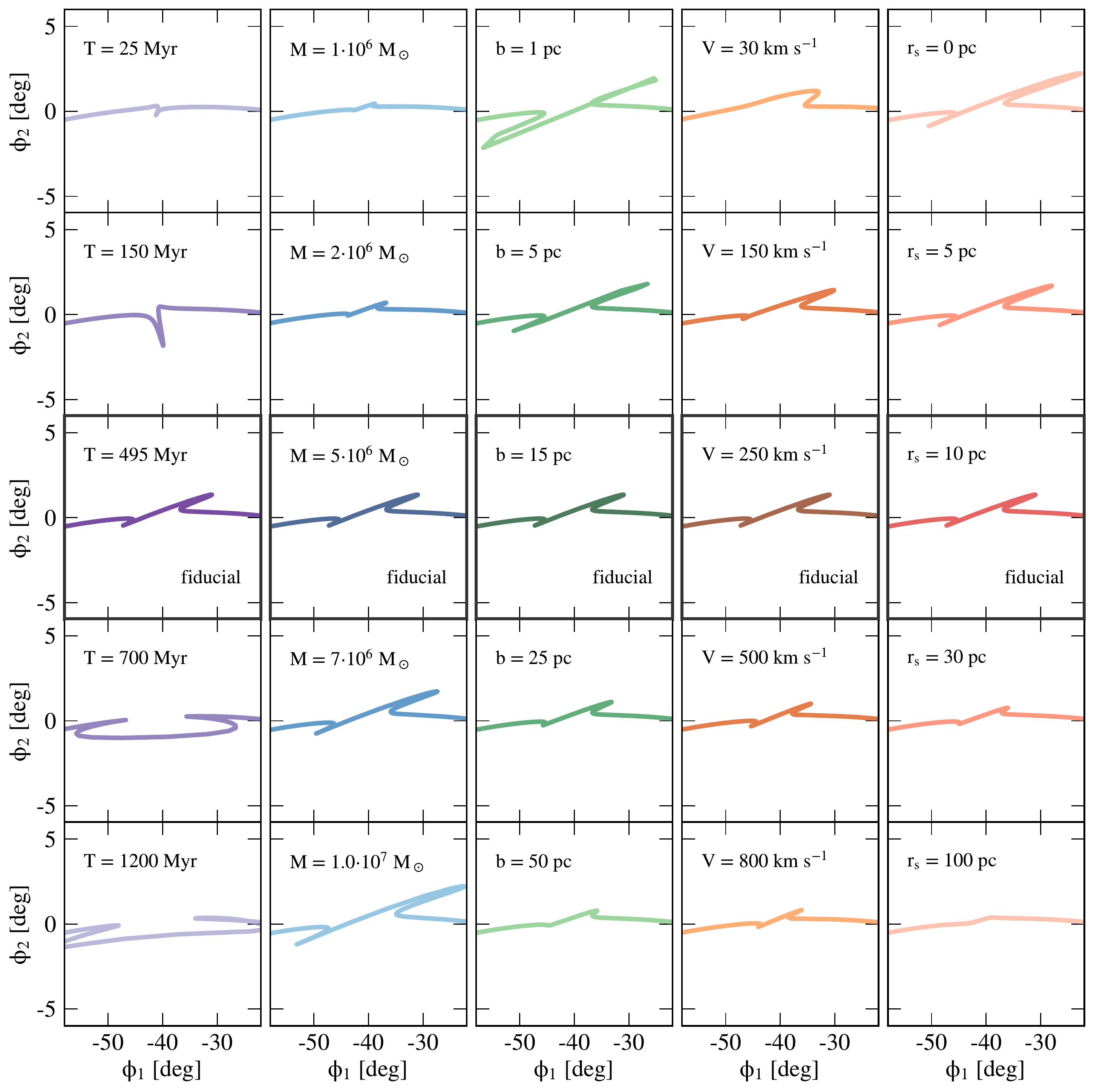}
\end{center}
\caption{The stream response to different encounter parameters in columns from left: the time of impact, perturber's mass, its impact parameter, velocity and size.
The value of the varied parameter (given in the top left of every panel) increases from top to bottom, while all the other parameters are at their fiducial values (given in the middle row).
Signatures of perturbation become more prominent with an increase in time after the impact and mass of the perturber, and with a decrease of its impact parameter, velocity and size.
The relative sizes of the gap and the loop, and the loop orientation, break some of these degeneracies.
}
\label{fig:scalings}
\end{figure*}

To efficiently explore the allowed parameter space, in this section we assume that GD-1 stream stars share the orbit with their progenitor.
This allows for even faster generation of a stream model, at the expense of less realistic density contrasts along the stream.
In the third row of Figure~\ref{fig:scalings} we show our fiducial model of perturbed GD-1 under this assumption.
The perturber parameters have the same normalization as the streakline model presented in the previous section, but the encounter velocity and impact parameter angles are different.
This modification was necessary because in general, stream tracks are offset from their progenitors' orbits \citep{sb2013}. 
Despite this difference, modeling the GD-1 stream with a single orbit results in features similar to those seen in the streakline model.
This simpler model also features a loop of stars removed from their original orbit, which projects to the GD-1 spur location at $\phi_1\gtrsim-40^\circ$, opens a gap at $\phi_1\approx-40^\circ$ and reconnects to the leading tail at $\phi_1\lesssim-40^\circ$.

With a method at hand to quickly generate stream models that reproduce basic features observed in GD-1, we explore how the stream morphology depends on impactor's properties.
We consider five parameters: time of impact, $T$, perturber's mass, $M$, its impact parameter, $b$, velocity, $V$, and size, $r_s$.
Each column of Figure~\ref{fig:scalings} shows models where one of the parameters is changed, while the others are kept at their fiducial values.
Parameter values are increasing from top to bottom (as labeled in top left of every panel), with the fiducial values shown in the middle row.
Most of the presented models preserve the angle between the loop and the unperturbed stream, while the main differences are in the width of the gap and the extent of the loop.

As the mass of the perturber increases, imparted velocity kicks to stream stars are larger, and the resulting loop and gap also increase in size.
The effects of other parameters are similar, for example, increasing the perturber's size (making it less dense) decreases the loop and gap sizes.
Indeed, the bottom right panel of Figure~\ref{fig:scalings} shows a model perturbed with an object following the $\Lambda$CDM concentration--mass relation \citep{diemer2018}, and it shows no signature of a loop.
Given time, both the loop and the gap grow in size.
However, older loops are more aligned with the stream (Figure~\ref{fig:scalings}, bottom left panel), and hence more difficult to detect observationally.
The presence of an observable spur alone already implies that GD-1 had a recent encounter with a dense perturber.

\begin{figure*}
\begin{center}
\includegraphics[width=\textwidth]{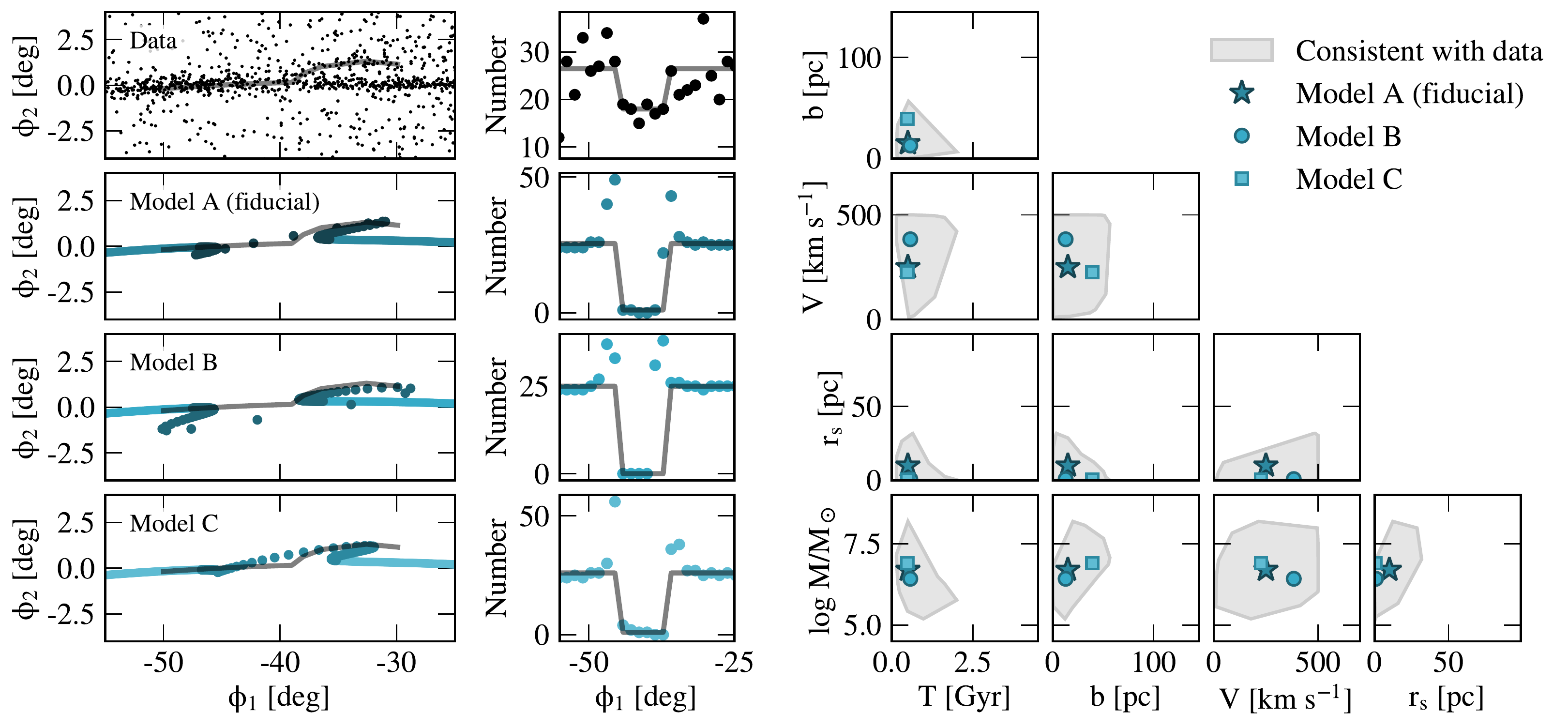}
\end{center}
\caption{(Left) Comparison of the observed GD-1 spur (left column) and gap (right column) in the top row to the modeled features in the bottom three rows. 
The gray line tracing the observed features was used for evaluating GD-1 models.
(Right) The parameter space of the allowed models is shaded gray, with the highlighted models falling inside this region.
The data prefer a recent, close encounter with a massive, dense object.
}
\label{fig:corner}
\end{figure*}

Decreasing the velocity of the perturber (which also increases the velocity kicks) produces larger loops and gaps, and this effect is nearly degenerate with an increase in perturber's mass at sufficiently large velocities.
However, at very low velocities, the whole loop morphology is changed (top panel in fourth column, Figure~\ref{fig:scalings}), likely because the encounter is no longer impulsive.
This means that the observed spur morphology cannot be explained by an object of an arbitrarily low mass moving slowly.
Likewise, decreasing the impact parameter has a unique signature too: the resulting loop is larger, but the gap size remains constant.
\changes{Different dependence of the loop and gap properties on the perturber's mass and its impact parameter suggests that both parameters can be constrained in the case of GD-1, where we observe both features.}

Different encounter parameters produce a unique impact on the GD-1 morphology, so we next search this space for parameter combinations allowed by the data.
Ideally, we would like to create a model of the stream and compare it to the data directly.
However, the adopted method for generating stream models is not sufficiently realistic for full forward-modeling.
Instead, we devised a set of criteria that allow us to compare whether a conceptual stream model represents the observed structure of GD-1.

First, we compare $\phi_2$ positions perpendicular to the stream of perturbed model stars to the position of observed stars in the spur, defined as a spline that goes through control points of the stream between $\phi_1=-50^\circ$ to $-39^\circ$, and control points on the spur between $\phi_1=-39^\circ$ to $-30^\circ$.
Positions of likely GD-1 members and this spline are shown in the top left of Figure~\ref{fig:corner} as black points and gray line, respectively.
Positions of stars in the fiducial model (A), and two other models (B and C) are shown in bottom rows, with darker colors marking stars with a significant change in energy due to the encounter (bracketed by the third percentile on the trailing end and the 92nd percentile of the energy difference on the leading end of the stream). 
Quantitatively, we define the spur chi-square:
\begin{equation}
\chi^2_\mathrm{spur} = \frac{1}{N}\sum_{i=1}^{N} \left(\frac{\phi_{2,\,i} - \phi_\mathrm{2,\,spline}(\phi_{1,\,i})}{\sigma_\mathrm{spur}}\right)^2 
\end{equation}
for $N$ model stars with positions $(\phi_{1,\,i}, \phi_{2,\,i})$, and adopted width of the spur as $\sigma_\mathrm{spur} = 0.2^\circ$.
To ensure that the model spur is long enough, we only consider models where at least one star has $\phi_1>-32^\circ$ and $\phi_2>0.8^\circ$.

Next, we compare the location and width of the observed and model gaps.
The observed gap profile is shown in the top panel of second column in Figure~\ref{fig:corner} (black points), and is well-represented by a top-hat profile centered on $\phi_\mathrm{1,\,gap}=-40.5^\circ$ and $w_\mathrm{gap}=8.5^\circ$ wide (gray line).
Gap profiles of models A through C have similar positions and widths, however, the density contrast between the gap and the spur is much larger (bottom rows).
As the predictions of our stream models regarding density are simplistic, we define the gap chi-square as:
\begin{equation}
\chi^2_\mathrm{gap} = \frac{1}{N_\mathrm{bin}}\sum_{i=1}^{N_\mathrm{bin}} \left(\frac{N_\mathrm{model}(\phi_{1,\,i}) - N_\mathrm{top}(\phi_{1,\,i})}{\sigma_\mathrm{gap,\,\mathit{i}}}\right)^2
\end{equation}
where $i$ denotes $N_\mathrm{bin}=29$ bins between $-60^\circ<\phi_1<-20^\circ$.
$N_\mathrm{model}(\phi_{1,\,i})$ is the number of model stars in bin $i$, and $\sigma_{gap,\,i}$ is the associated Poisson uncertainty.
$N_\mathrm{top}(\phi_{1,\,i})\equiv N_\mathrm{top}(n_\mathrm{base}, n_\mathrm{hat}, \phi_\mathrm{1,\, gap}, w_\mathrm{gap} | \phi_{1,\,i})$ is the number of stars expected in bin $i$ from a top-hat distribution with the position $\phi_\mathrm{1,\,gap}$ and width $w_\mathrm{gap}$ adopted from the observed profile, but with the base level, $n_{\mathrm base}$, given by the median of the model profile outside of the gap ($-55^\circ<\phi_1<-45^\circ$ and $-35^\circ<\phi_1<-25^\circ$), and $n_\mathrm{hat}$ is the median of the model profile in the gap ($-43^\circ<\phi_1<-37^\circ$).
We further require the density contrast $n_\mathrm{hat} / n_\mathrm{base}$ at least 0.5.

We combine the spur and gap contributions to the surrogate log likelihood as $\ln\mathcal{L} = -(\chi^2_\mathrm{spur} + \chi^2_\mathrm{gap})$.
Even though this likelihood is an approximation to the formal likelihood, which would compare the positions of model stream stars to the observed ones, it is expected to favor models that reproduce well features seen in the data (spur and gap).
Therefore, we use an ensemble MCMC sampler \citep{Foreman-Mackey:2013} to find the allowed range in parameters of interest: perturber's mass, size, velocity, impact parameter and impact time, while marginalizing over the orientation angles and impact location.
We started 200 walkers and advanced them for 5000 steps, keeping the last 2000 steps for analysis.
In principle, the resulting chain provides posterior samples, but since this is a highly idealized search of the parameter space, we only provide plausible ranges of parameters, instead of showing their two-dimensional distributions.
Specifically, in each panel of the corner plot (Figure~\ref{fig:corner}, right) we show the two-dimensional convex hull of all models with likelihood above the 5th percentile (gray shaded regions).
Therefore, Figure~\ref{fig:corner} contains likelihood, rather than posterior, information.
\changes{Applied to the fiducial model of GD-1 (presented in Figure~\ref{fig:fiducial}), this method recovers the true encounter parameters.}

High-likelihood models of the GD-1 stream have encounter parameters expected from the stream's sensitivity to different parameters (explored in Figure~\ref{fig:scalings}).
Recent encounters are favored, with most of the models having been perturbed within the last 1\,Gyr and none more than 2\,Gyr ago.
The perturbing object itself is massive ($5.5\lesssim\log M/\rm M_\odot\lesssim8$) and dense ($r_s\lesssim20\,\rm pc$).
A range of velocities are allowed, as the data dismiss only the slowest moving perturbers ($V\gtrsim50\,\rm km\,s^{-1}$).
The closest approach was extremely close to the stream, with the impact parameter smaller than $b\lesssim50\rm\,pc$.

\changes{Many of the inferred encounter parameters are correlated.
For example, more massive perturbers allow for larger impact parameters, larger scale radii and older encounters.
In addition, current data places very weak constraints on the perturber's velocity.
This is hardly surprising given the similar effect these parameters have on the appearance of the spur and the gap (Figure~\ref{fig:scalings}).
In the next section we discuss new observables that can further constrain the GD-1 perturber.
}

\subsection{Predictions for the kinematic signatures of the encounter in GD-1}
\label{sec:kinematics}
Perturber properties presented in the previous section were constrained by the spatial distribution of the GD-1 debris alone.
We now explore kinematic signatures of the GD-1 encounter with such a perturber and discuss how to further constrain its properties with future kinematic observations.

\begin{figure}
\begin{center}
\includegraphics[width=\columnwidth]{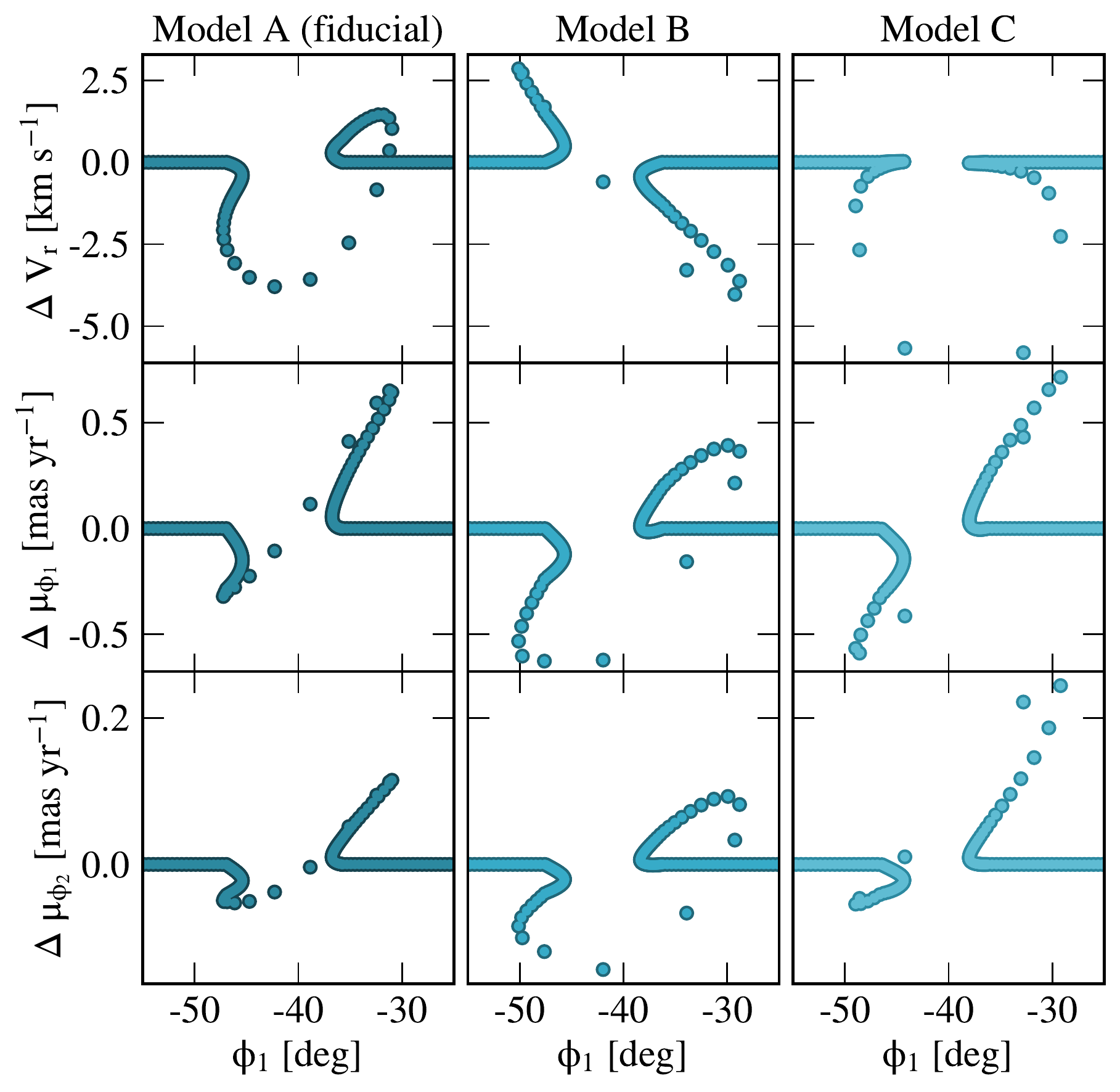}
\end{center}
\caption{Components of the relative velocity between the perturbed and unperturbed stream (with radial velocity on top, and proper motion along and perpendicular to the stream in the middle and bottom, respectively) for three different model streams (columns).
The direction of the offset in radial velocity depends on the encounter geometry and can constrain the perturber's orbit.
The direction of velocity offsets in proper motions is universal and can be used to falsify the encounter scenario.
}
\label{fig:predictions}
\end{figure}

In Figure~\ref{fig:corner} we introduced three models whose loops reproduce the observed spur and gap morphologies (left), and consequently, their encounter parameters are in the allowed region of the parameter space (right).
These particular models were selected to display the diversity in kinematic signatures of the loop allowed by the spatial data.
To account for the velocity gradients present along the stream, in Figure~\ref{fig:predictions} we show differences in velocity components between the perturbed and unperturbed stream for these three models.
Each column displays velocity signatures of a single model, with radial velocity differences in the top row, and the two proper motion components in GD-1 coordinates in the following rows.
In all of the models, there are kinematic offsets between the loop and the unperturbed stream.

Radial velocity signatures exhibit the largest diversity, most prominently manifested as the velocity difference between the spur and the unperturbed stream at $\phi_1\approx-30^\circ$.
Here, model A predicts the spur to move faster than the stream, in model B the stream moves faster than the spur, whereas there is no difference between the two in model C.
The other side of the loop has the opposite behavior, so models A and B predict bimodal radial-velocity measurements in the GD-1 stream at $-50^\circ\lesssim\phi_1\lesssim-45^\circ$.
Models A, B, and C have rather similar parameters overall (Figure~\ref{fig:corner}, right), but the perturber's orbital plane is misaligned with that of GD-1 by $170^\circ$, $15^\circ$, and $50^\circ$, respectively.
Therefore, future follow-up observations will not only further constrain all impact parameters, but also provide the first strong constraints on the encounter geometry.

Velocity offsets in proper motions have the same direction across all models (Figure~\ref{fig:predictions}, bottom two rows).
Focusing first on the $\phi_1$ direction, this behavior stems from the fact that there is a gap in the stream and that the stream moves in the negative $\phi_1$ direction (i.e., $\mu_{\phi_1}<0$).
As \citet{eb2015} showed, velocity kick along the stream (the $\phi_1$ direction) makes the affected stars on the leading side of the gap speed past the unperturbed stream.
And indeed, loop stars at $\phi_1\lesssim-45^\circ$ have proper motion $\mu_{\phi_1}$ more negative than the unperturbed stream at the same $\phi_1$.
The opposite behavior is expected, and seen, on the other side of the loop ($\phi_1\gtrsim-35^\circ$), where the affected stars lag behind the stream at less negative $\mu_{\phi_1}$.
Details of the loop proper motion offsets still depend on the perturber's parameters, but the direction of offsets along the stream is universal in the encounter scenario.

Because the spur is at positive $\phi_2$, its proper motion perpendicular to the stream (the $\phi_2$ direction) should be faster than that of the unperturbed stream stars.
As expected, $\Delta\mu_{\phi_2}>0$ for all of the models at $\phi_1\gtrsim-35^\circ$.
The other side of the loop is at most slightly offset from the main stream track, so $\Delta\mu_{\phi_2}$ is very small at $\phi_1\lesssim-45^\circ$.
Finally, the magnitude of kinematic offsets perpendicular to the stream ($\Delta\mu_{\phi_2}$) is universally smaller than along the stream ($\Delta\mu_{\phi_1}$).
This is also expected because the gap is much wider than the distance between the stream and the spur.

The encounter model makes falsifiable predictions for proper motion kinematics in the affected region of the GD-1 stream: the spur, as a trailing part of the loop, should be moving slower than the adjacent stream stars in the $\phi_1$ direction (along the stream).
The predicted velocity offsets are small, but proper motions of diffuse, cold streams have been measured at this precision with the \hst\ \citep{sohn2016}, so future kinematic data will rule on the perturbative origin of GD-1 features.
Furthermore, the expected offsets in radial velocity are on the order of a few km\,s$^{-1}$, which is easily within reach of current spectroscopic facilities \citep[e.g.,][]{sg2007}.
The baseline radial velocity in the perturbed part of the stream is $\sim-100\,\rm km\,s^{-1}$ \citep{koposov2010}, allowing for confident selection of GD-1 members out of the field Milky Way population.
Should the encounter scenario be confirmed, these new data will make strong predictions on the orbit and present-day position of the pertuber.

\section{Discussion}
\label{sec:discussion}
In previous section we presented a model of the GD-1 stream which experienced a recent encounter with a massive, dense object.
This fly-by imparted significant velocity kicks to the closest stars both along the stream, which produced a gap in the stream, and perpendicular to the stream, which launched a spur of stars outside of the main stream.
The qualitative and quantitative agreement between the observed and modeled gap and spur suggests that these GD-1 structures are the first dynamical evidence for \changes{a} halo substructure.
Below we discuss improvements for future modeling of GD-1 (Section~\ref{sec:caveats}), review possible origin scenarios of the observed features (Section~\ref{sec:origin}), and outline strategies to distinguish these scenarios with additional observations (Section~\ref{sec:future}).

\subsection{Limitations of the current modeling setup}
\label{sec:caveats}
A number of simplifying assumptions were made to facilitate the initial modeling of complex features observed in the GD-1 stream.
While we expect these assumptions to still produce unbiased inference, there is additional information available in the present data that has yet to be employed.
We will need better models to fully explain these observations, so in this section we discuss areas for improvement in modeling of the GD-1 system.

Our fiducial model of GD-1 is built from test particles released by the disrupting progenitor directly from the Lagrange points, instead of particles dynamically ejected from the globular cluster progenitor itself.
Stream models generated under this assumption match the distribution of tidal debris (including the intrinsic density variations along the tidal tails) from direct N-body simulations while the progenitor persists \citep[e.g.,][]{kupper2012,fardal2015}.
However, the method is yet to be tested when the progenitor is completely disrupted, such as in the case of GD-1, and most of the known halo streams.
The last stages of tidal disruption can result in enhanced density variations, so to fully account for all the structures observed in GD-1 we will need a full self-gravitating N-body model of the stream.

Of course, the reason we decided against employing self-gravitating models in this work is that they are computationally expensive, and prohibitively so for any kind of parameter space exploration.
This is why we further focused only on the perturbed region of GD-1 when inferring properties of its perturber, and assumed that all stars are on the same orbit.
Reassuringly, fiducial encounter parameters produce qualitatively similar features both in an idealized stream (Figure~\ref{fig:fiducial}) and in stars along a single orbit (Figure~\ref{fig:scalings}).
As discussed in Section~\ref{sec:perturber_properties}, this choice limited us to only matching positions of stream features, and to disregard density along the stream.
Since the density profile of the gap is also expected to contain information about the perturber \citep{eb2015b}, going forward we will need to have a proper generative model of the stream.

Furthermore, our treatment of the GD-1 perturber is also simplistic.
Although we explored perturbers of different masses and sizes, they all follow the \citet{hernquist1990} density profile.
This profile reduces to the point mass case in the limit of vanishing scale size, and is similar to the profile of dark matter halos \citep[NFW,][]{navarro1997} at small radii, although at larger radii it falls off more steeply as $r^{-4}$, compared to $r^{-3}$ for NFW halos.
While the inference of the perturber's scale radius should be robust to the details of its outer density profile, future work should explore whether any observables can be traced to the perturber's density structure, as it might hold additional clues to its origin.

In the absence of a realistic stream model, our inference of the GD-1 perturber was based on a set of high-level criteria instead of directly calculating the likelihood of the observed spatial distribution of tidal debris for a given set of model parameters \citep[e.g., likelihood developed in][]{bonaca2014}.
Spot-checking of the accepted models suggests we erred on the conservative side and accepted a wide range of models, some of which would likely be ruled out with the full likelihood.
For example, model B in Figure~\ref{fig:corner} has both sides of its loop appreciably offset from the stream track, instead of just the trailing side coinciding with the spur.
Because of that, with the current approach we only bound the allowed parameter space, and remain agnostic to the relative likelihood of models within the bounds.
\changes{This prohibits us from measuring probabilities associated with each model parameter.
For example, in the following section we qualitatively discuss the inferred constraints on the mass and size of the GD-1 perturber in the context of objects orbiting the Milky Way, but are unable to quantify their relative probabilities.}
To find preferred regions of the parameter space and recover the posterior density, future inference will need to employ a more realistic model of GD-1 and a proper likelihood.

And finally, our search of the parameter space has not been exhaustive, so different islands may still be allowed (for example, at a lower perturber mass or older encounter).
This is, of course, always true when sampling a parameter space \citep[e.g.,][]{hoggdfm2018}.
However, ours is sufficiently low-dimensional that future studies employing the formal likelihood should be able to perform a brute-force sweep of the entire parameter volume and identify all classes of perturber properties that explain GD-1 features.

\subsection{Origin of GD-1 structures}
\label{sec:origin}
Many massive objects are known, or expected, to orbit the Galaxy, and if any should have encountered a cold stream, the perturbation would remain recorded in the stream's density profile.
So, the presence of perturbed features in GD-1 is not surprising, but their detailed structure is.
In this section we discuss what the observed structures suggest about their origin.

\subsubsection{Non-impact scenarios}
In this work we assumed that GD-1 was perturbed because of \changes{inhomogeneities} in its density profile.
However, N-body simulations of disrupting globular clusters show that a certain level of structure in the resulting stream is intrinsic to the disruption process \citep[e.g.,][]{kupper2008,kupper2010}.
\changes{These simulations predict that stars escape a globular cluster on orbits that epicycle around the progenitor's orbit.
Seen in projection, these epicycles produce a regular pattern of over- and under-densities along the stream.
\citet{kupper2015} showed that some of density variations detected in tidal tails of the Palomar~5 globular cluster can be attributed to epicycles.
}
\changes{However,} none of the present-day globular clusters would \changes{intrinsically} produce density variations as prominent as those observed in GD-1 \citep[see][]{pwb}.
If the GD-1 progenitor had a more complex internal structure than the known globular clusters, this might propagate to the morphology of its tidal tails and account for the observed features.

Non-trivial morphologies in stellar streams can also be produced by chaotic dispersal \citep{pw2016}.
For example, unexpectedly short lengths of the Ophiuchus and Palomar~5 streams have been attributed to chaos: in a chaotic potential featuring a rotating bar, large swaths of these streams can be dispersed to surface densities below our current detection threshold \citep{pw2016b,pearson2017}.
GD-1, on a retrograde orbit and with a larger pericenter than either of these streams, is deemed less susceptible to the chaotic influence of the rotating bar \citep{bb2018}.
Furthermore, the gap and spur features in GD-1 are much more localized than typical signatures of chaos in complex, time-dependent gravitational potentials, but there is still a lot of parameter space left to explore.

\begin{figure}
\begin{center}
\includegraphics[width=\columnwidth]{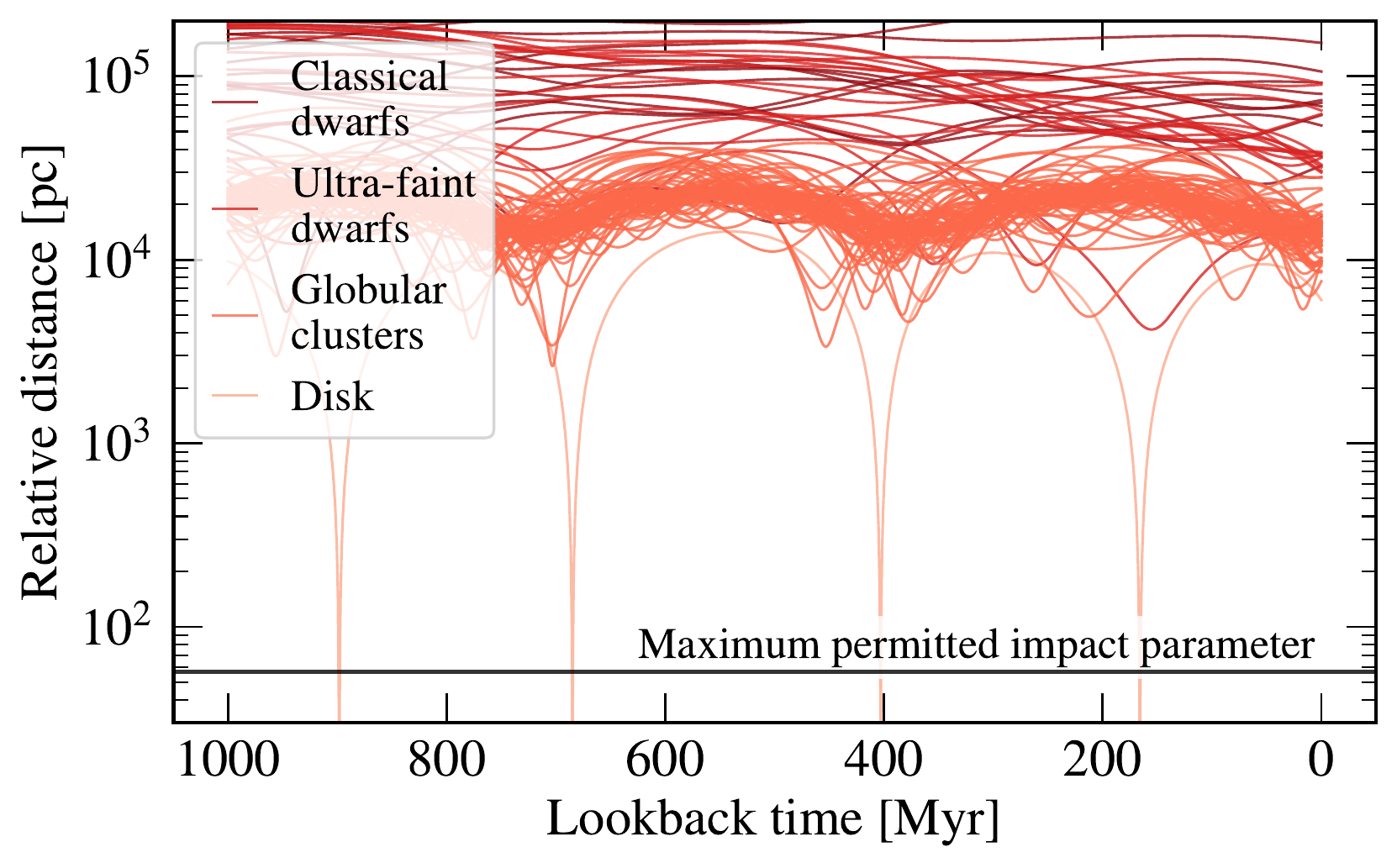}
\end{center}
\caption{Relative distance between the GD-1 gap and known objects in the Milky Way: classical dwarf galaxies (dark red), ultra-faint dwarfs (red), globular clusters (orange) and the stellar disk (light orange).
The horizontal line shows the maximum permitted impact parameter, as shown in Figure~\ref{fig:corner}.
No known, compact object approaches GD-1 close enough to produce the observed gap-and-spur features.
}
\label{fig:known_encounters}
\end{figure}

\begin{figure*}
\begin{center}
\includegraphics[width=0.9\textwidth]{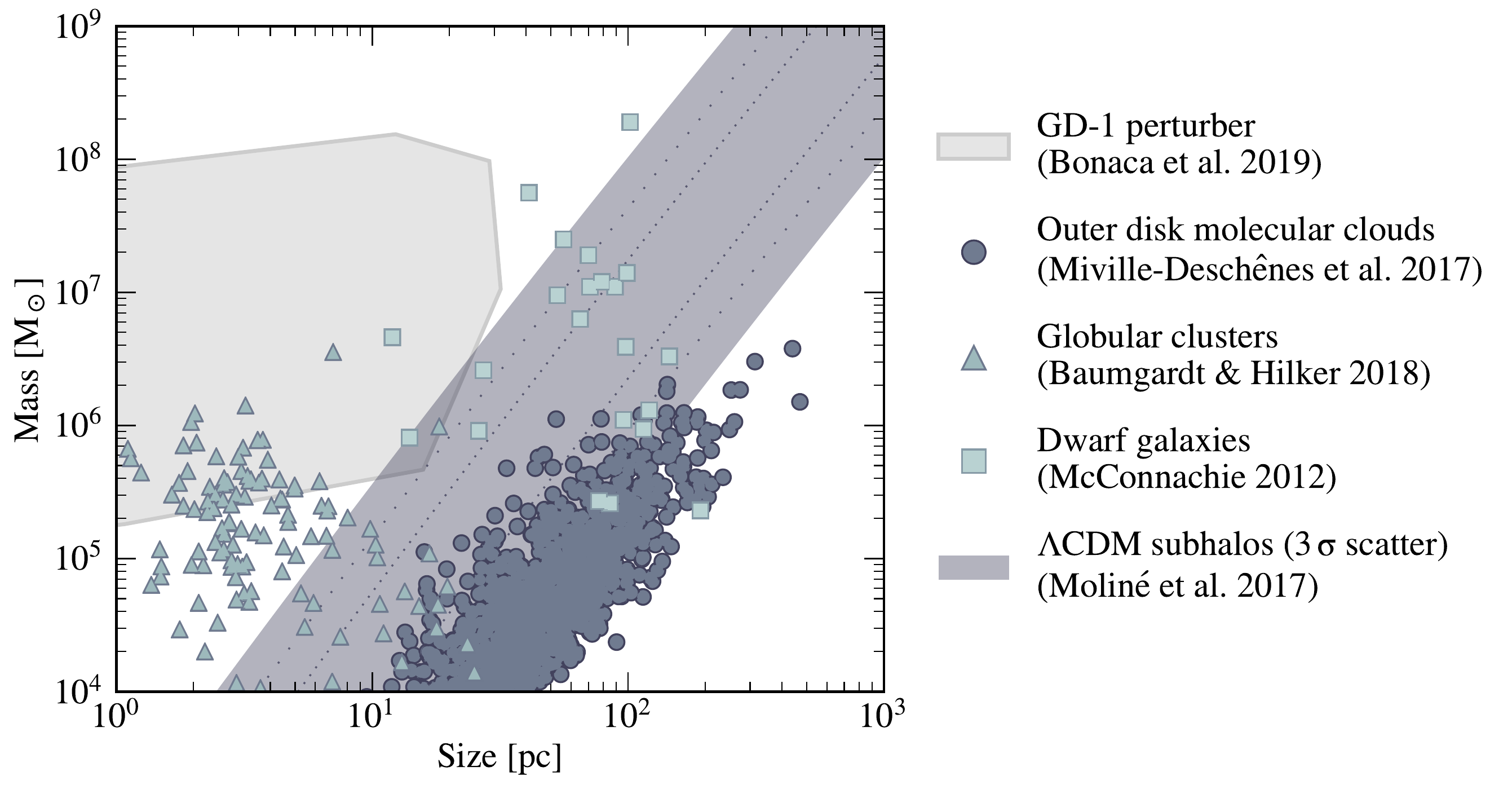}
\end{center}
\caption{Comparison of inferred mass and scale radius of the GD-1 perturber (following a Hernquist profile, light gray shaded region) to the known dwarf galaxies (squares), globular clusters (triangles), and molecular clouds in the outer disk (circles).
For dwarf galaxies and globular clusters we show the total mass and half-light radius, while for molecular clouds we show total mass and total size.
Molecular clouds are too diffuse to have caused features in the GD-1 stream, while orbital properties rule out globular clusters and dwarf galaxies.
The dark shaded region is showing masses and scale radii of dark matter subhalos (following an NFW profile) and the expected $3\,\sigma$ scatter (the inner and outer dotted lines denote the $1\,\sigma$ and $2\,\sigma$ scatter, respectively).
GD-1 perturber is on the dense, or high-concentration, end of dark matter subhalos.
}
\label{fig:mass_size}
\end{figure*}

\break
\subsubsection{Luminous objects as the GD-1 perturber}

While these non-interaction scenarios are still possible causes of structures seen in GD-1, we find the interaction scenario more plausible.
The Milky Way is surrounded by $\sim50$ dwarf galaxies \citep{mcconnachie2012} and $\sim150$ globular clusters \citep{harris2010}, most of which, like GD-1, reside in the Galactic halo.
Thanks to the \gaia\ mission, a significant fraction of these satellites are now fully located in the 6D phase space \citep{simon2018, gdr2_satellites, baumgardt2019}.
To test whether any of these objects could have perturbed GD-1, we integrated their orbits in a fiducial Milky Way potential \citep{pw2017} and show their relative distance from the GD-1 gap at $\phi_1=-40^\circ$ during the past billion years in Figure~\ref{fig:known_encounters}.
Lines are shaded according to the object's mass, with the classical dwarfs being the darkest, ultra-faints medium, and globular clusters lighter.
All of them have kept at least 1\,kpc away from GD-1.
Relative distances shown in Figure~\ref{fig:known_encounters} were calculated for fiducial present-day positions and velocities of satellites.
Since the associated measurement uncertainties can be substantial, we also sampled the error distribution in satellite properties and measured in how many realizations any given satellite comes closer to GD-1 than the maximum permitted impact parameter of 57\,pc (see Figure~\ref{fig:corner}).
Upon resampling the observational uncertainties, all satellites \changes{with known orbits} are excluded as GD-1 perturbers at a $3\,\sigma$ level or higher. 
\changes{Orbital parameters are known for all but two globular clusters \citep[GLIMPSE C02 and 2MASS-GC01,][]{baumgardt2019}, both of which are faint objects located in the Galactic bulge \citep{kurtev2008, bonatto2008}, and are therefore unlikely GD-1 perturbers.
Out of 49 dwarf galaxies present in the up-to-date catalog of \citet{mcconnachie2012}, only 24 objects have their orbits determined.
In summary, present data rule out all luminous satellites with known orbits as GD-1 perturbers, but follow-up spectroscopy is required to test the remaining 25 satellites.
}

As it orbits the Galaxy, GD-1 crosses the disk at timescales comparable to the inferred time of perturbation (the lightest line in Figure~\ref{fig:known_encounters} is the distance from the Galactic plane).
While strong disk shocking can facilitate disruption of a diffuse globular cluster \citep{dehnen2004}, GD-1 disk crossings are between 13\,kpc and 23\,kpc from the Galactic center, where the disk density is too low to significantly impact the stream, or produce sharp features such as the gap and the spur.
Still, giant molecular clouds (GMCs) that are orbiting in the disk plane can perturb cold stellar streams \citep{amorisco2016}.
To test whether GMCs are viable candidates for the GD-1 perturber, in Figure~\ref{fig:mass_size} we compare the inferred mass and size of the GD-1 perturber (gray shaded region) to known objects in the Milky Way, including molecular clouds.
Dwarf galaxies are shown as light squares \citep{mcconnachie2012}, globular clusters as medium triangles \citep{baumgardt2018}, and outer-disk molecular clouds (beyond 10\,kpc) as dark circles \citep{md2017}.
This comparison is rather conceptual as different classes of objects have different density profiles: for the GD-1 perturber we show the mass and scale radius assuming a Hernquist profile, for dwarf galaxies and globular clusters we show the total dynamical mass and half-light radius, and the total (high) mass and full size for molecular clouds.
Keeping these caveats in mind, the most massive globular clusters and the most compact dwarf galaxies have masses and sizes comparable to the preferred values of the GD-1 perturber, but GMCs are in general too diffuse \changes{(at a given mass, their sizes are at least an order of magnitude larger than expected of the GD-1 perturber)}.
To additionally test for extremely dense, yet undiscovered class of GMCs, we also created GD-1 models perturbed by a $10^7\,\rm M_\odot$ point mass moving on a circular disk orbit for the three most recent disk crossing times.
These configurations result in a spur that is below the stream at $\phi_2<0$, rather than above at $\phi_2>0$ as observed in GD-1.
\changes{Based on their low central density and their expected orbits, we conclude that GMCs are unlikely to have perturbed GD-1.
A dense GMC orbiting outside of the Milky Way disk at large Galactocentric radii is still allowed by the data, and in Section~\ref{sec:future} we discuss future tests to ascertain the nature of the GD-1 perturber.
}

\changes{
The prospect of ruling out known luminous objects as GD-1 perturbers based on their orbits strongly depends on the accurate knowledge of the underlying gravitational potential.
In this work we assumed a smooth and static model for the Milky Way \citep{pw2017} because it reproduces well the global 6D phase-space distribution of GD-1 \citep{pwb}.
This implies that, to first order, the assumed gravitational potential is close to the effective Milky Way potential over the last 3\,Gyr (the dynamical age of GD-1 in our fiducial model).
Future studies will explore whether GD-1 can distinguish between different choices for a static potential \citep[e.g.,][]{bovy2015}.
Furthermore, the presence of two massive satellites within the Milky Way, Large Magellanic Cloud (LMC) and Sagittarius, means the potential is more complex in detail.
Dynamical considerations imply that the LMC is very massive \citep[$\sim10^{11}\rm M_\odot$, e.g.,][]{kallivayalil2013, gomez2015, penarrubia2016}, and has already been invoked to explain deviations from the expected stream tracks of the Tucana~III and Orphan streams \citep[respectively]{erkal2018, erkal2019}.
Sagittarius is likely less massive \citep[$\sim10^9-10^{11}\rm M_\odot$,][]{jiang2000}, but still a significant perturber in the inner Galaxy \citep[e.g.,][]{laporte2019a, laporte2019b}.
These massive satellites may have affected the orbit of GD-1, as well as those of luminous satellites and any dark-matter subhalos.
To reaffirm that the GD-1 perturber is not a luminous satellite, future work should recalculate their relative distances in a more realistic gravitational potential that includes the LMC and Sagittarius \citep[for example, as a time-dependent expansion of basis functions,][]{ngc2019}.
}

\changes{
So far, we have estimated the influence of known objects on GD-1 by directly modeling individual bodies.
An alternative approach may be to consider a statistical model for the distribution function of GMCs, globular clusters and dwarf galaxies in our Galaxy.
For example, \citet{banik2019} compared the power-spectra of gaps in the Palomar~5 stream expected from whole populations of GMCs and dark-matter subhalos.
A similar analysis of GD-1 could, in principle, estimate the likelihood of the GD-1 perturber being a member of these groups.
However, unlike the gaps in Palomar~5, the gap detected in GD-1 is associated with an off-stream spur.
A statistical treatment of both features is beyond the scope of this work, but the framework for simultaneous modeling of a gap-and-spur feature that we developed here should provide a good starting point for such population studies in the future.
}

\subsubsection{Dark perturbers}
Having ruled out \changes{luminous objects with known orbits}, we find that the most probable GD-1 perturber is a dark object in the Milky Way halo.
As luminous satellites can have approximately the required masses and sizes, a low luminosity unknown satellite might be the culprit.
To avoid detection, it would have to be fortuitously hiding in the disk plane, or moving very fast, as our best estimate is that the encounter was recent.
Near-future and upcoming surveys of the plane \citep{schlafly2018} and the halo \citep{lsst} should provide a complete census of objects in the Milky Way and find a perturber of low luminosity.

Alternatively, the GD-1 perturber could be completely dark.
A dense pertuber in the mass range $10^5-10^8\,\rm M_\odot$ is required, so we next discuss black holes -- the densest dark objects in the Universe.
Baryonic black holes of similar masses typically reside in centers of galaxies \citep[the mass of Milky Way's supermassive black hole, Sgr A$^\star$, is $\approx4\times10^6\,\rm M_\odot$][]{boehle2016}.
A population of non-baryonic, primordial black holes is hypothesized to have formed in the early universe \citep{carr1974}, and has sparked a renewed interest as a dark matter candidate following the LIGO detections \citep{bird2016}.
Several lines of inquiry have limited the contribution of massive ($\gtrsim10^3\,\rm M_\odot$) primordial black holes to the dark matter budget to less than $\lesssim0.1\,\%$ \citep[and references within]{carr2016}.
So if GD-1 encountered a primordial black hole, this would have been an extremely rare event, and we would not expect to see similar features in other streams upon a comparable amount of scrutiny.

On the other hand, $\Lambda$CDM cosmological simulations predict scores of dark matter subhalos orbiting Milky Way-like galaxies.
Even after accounting for the destruction of subhalos due to the stellar disk \citep{donghia2010,errani2017,gk2017}, their density in the inner 20\,kpc is high enough that a stream on a GD-1-like orbit is expected to have encountered at least one $10^6-10^7\rm\, M_\odot$ subhalo within the last $\sim8$\,Gyr \citep{erkal2016}.
Of the two prominent gaps in GD-1, our fiducial model ascribes one to the site of the progenitor's disruption (G-20), and one to the encounter with a perturber (G-40).
Thus, a dark matter subhalo is a plausible GD-1 perturber in terms of encounter rates expected in the $\Lambda$CDM universe.

The high inferred density of the GD-1 perturber makes it more resilient to disruption in the tidal field of the inner galaxy, but preferred values are on the high end of dark matter halo concentrations.
For example, the \lcdm\ concentration--mass relation for isolated dark matter halos predicts that a $10^6\,\rm M_\odot$ halo should have a scale radius of $\sim50\,\rm pc$ \citep{diemer2018}, while the most diffuse GD-1 perturber of a similar mass has a scale radius of $\sim20\,\rm pc$.
The scatter in the mass-concentration relation is small at high masses ($\sim0.15\,\rm dex$), although it has not been quantified below $\lesssim10^{10}\,\rm M_\odot$ and there are some indications that the fraction of high-concentration halos increases at low masses \citep{diemer2015}.
On the other hand, subhalos surviving in dense environments are more concentrated than field halos of the same mass \citep[e.g.,][]{avilareese2005}.
The dark shaded band in Figure~\ref{fig:mass_size} shows the masses ($M_{200,c}$) and sizes (NFW scale radii, $r_s$) expected for dark matter subhalos orbiting in the GD-1 radial range \citep{moline2017}.
The whole band encompasses the $3\,\sigma$ scatter in the concentration--mass relation, while the dense and sparse dotted lines correspond to the $1\,\sigma$ and $2\,\sigma$ scatter, respectively.
Properties of \lcdm\ subhalos are consistent with the inferred mass and size of the GD-1 perturber \changes{only} at a $2-3\,\sigma$ level, \changes{and the} current GD-1 analysis allows the perturber to be orders of magnitude denser than \changes{expected of $\Lambda$CDM} dark matter subhalos.

\changes{Among the objects expected to orbit the Milky Way, globular clusters have the largest overlap with the inferred mass and size of the GD-1 perturber (Figure~\ref{fig:mass_size}).
Current dynamical data rule out all of the known globular clusters, but this overlap motivates a search for new globular clusters, which we discuss in the next section.
Should future searches for luminous objects yield no plausible candidate for the GD-1 perturber, and a dark-matter subhalo remains a viable option, the high inferred density might point to dark matter physics beyond CDM \citep[e.g.,][]{kahlhofer2019}.
}

\subsection{Future prospects}
\label{sec:future}
Constraining the number and properties of low-mass dark matter halos is essential for understanding the nature of dark matter \citep{bullockmbk2017}.
Interpreting the gap-and-spur feature in the GD-1 stream as a signature of a perturbation provides us a really promising candidate for a dark halo substructure.
With the follow-up work outlined below, we now have a great opportunity to detect and characterize a \emph{dark} dark matter subhalo.

The best way to test whether GD-1 features are an outcome of an encounter with a massive perturber is to obtain detailed kinematics in the perturbed region of the stream.
The encounter scenario predicts velocity offsets between the spur and the stream, and they can be used to falsify this model.
The expected offsets are small, but measurable with a modest investment of spectroscopic time to get the radial component of the velocity (the required precision is $\sim1\,\rm km\,s^{-1}$), and a somewhat more significant astrometric commitment to measure the transverse motion precisely enough ($\sim0.1\,\rm mas\,yr^{-1}$).
The magnitude of these velocity offsets could already falsify other origin scenarios.
For example, a velocity offset between the spur and the stream that is much larger than the velocity dispersion in a globular cluster would imply an input of energy to the system, and thus rule out scenarios in which the spur is a consequence of substructure in the GD-1 progenitor.
However, the ultimate test of the encounter scenario is a measurement of velocity offsets on both sides of the GD-1 gap.
The offsets are predicted to monotonically increase along the loop, with the spur always lagging behind the stream in the component of the velocity along the stream (for details, see Section~\ref{sec:kinematics}).

Should the encounter scenario be confirmed, details of the kinematic structure in the perturbed part of GD-1 will put very strong constraints on the orbit of the perturber (see Figure~\ref{fig:predictions} for different possibilities).
This would in turn allow the search for additional signatures of the perturber along its inferred orbit.
Recently, \citet{vantilburg2018} showed how a massive and dense object can be detected in the halo via time-dependent weak lensing of background stars.
As the magnitude of the effect depends on the object's density profile, we might directly measure structural properties of the GD-1 perturber.
Combined with a better theoretical understanding of dark matter subhalos and their density profiles, this measurement could also inform about the particle nature of dark matter.

Ultimately, locating the perturber gravitationally would open up possibilities for electromagnetic follow-up.
If the source is bright in x-rays, this might be a signature of an accreting black hole \citep[e.g.,][]{bailyn1995}.
Alternatively, a spatially coincident excess of gamma-rays might signal the annihilation of dark matter particles \citep[similarly to the results of searches at the locations of dwarf galaxies, e.g.,][]{hooper2015}.
Either way, an electromagnetic detection would better characterize the nature of the GD-1 perturber.

As discussed above, further study of the GD-1 gap-and-spur feature may provide the first opportunity to follow up an individual halo substructure.
But most excitingly, these features demonstrate that cold stellar streams are extremely fine-tuned detectors, sensitive at a level that was only hoped for beforehand.
\changes{In GD-1 alone, there are additional gaps that may be further evidence of gravitational perturbations (specifically, the G-20 gap is associated with a diffuse blob of GD-1 stars beyond the main stream).}
In addition to GD-1, there are over 40 known streams in the Milky Way halo \citep[e.g.,][]{gc2016, shipp2018}.
In the era of \gaia, we now have both the incentive and the resources to study them all in detail.
With the full network of streams we could learn not only about individual halo substructures, but about the population as a whole.

\acknowledgements
We thank the referee for thorough comments that improved the manuscript.
It is also a pleasure to thank Vasily Belokurov, Warren Brown, Will Dawson, Benedikt Diemer, Elena D'Onghia, Adrienne Erickcek, Douglas Finkbeiner, Lars Hernquist, Kathryn Johnston, Manoj Kaplinghat, Sergey Koposov, Doug Lin, Barry McKernan, Erica Nelson, Sarah Pearson, Hans-Walter Rix, Josh Speagle, and Tomer Yavetz for valuable discussions and input.

This project was developed in part at the 2018 \acronym{NYC} \Gaia\ \acronym{DR2} Workshop at the Center for Computational Astrophysics of the Flatiron Institute in New York City in 2018 April.

This work was performed in part at Aspen Center for Physics, which is supported by National Science Foundation grant PHY-1607611.

This work has made use of data from the European Space Agency (\acronym{ESA}) mission \Gaia\ (\url{https://www.cosmos.esa.int/gaia}), processed by the \Gaia\ Data Processing and Analysis Consortium (\acronym{DPAC}, \url{https://www.cosmos.esa.int/web/gaia/dpac/consortium}). Funding for the \acronym{DPAC} has been provided by national institutions, in particular the institutions participating in the \Gaia\ Multilateral Agreement.

\software{
\package{Astropy} \citep{astropy:2013, astropy:2018},
\package{colossus} \citep{diemer2018},
\package{gala} \citep{pw2017},
\package{emcee} \citep{Foreman-Mackey:2013},
\package{IPython} \citep{Perez:2007},
\package{matplotlib} \citep{Hunter:2007},
\package{numpy} \citep{Van-der-Walt:2011},
\package{scipy} \citep{scipy}
}

\bibliographystyle{aasjournal}
\bibliography{spur}

\end{document}